\documentclass[a4paper]{jpconf}
\usepackage{graphicx}
\newcommand{\jpsi}{\ensuremath{\mathrm{J/}\psi}} 
\newcommand{\snn}{\ensuremath{\sqrt{s_\mathrm{NN}}}}
\newcommand{\tev}{\ensuremath{\,\mathrm{TeV}}}
\newcommand{\gev}{\ensuremath{\,\mathrm{GeV}}}
\newcommand{\pT}{\ensuremath{ p_\mathrm{T}}}
\newcommand{\raa}{\ensuremath{ R_\mathrm{AA}}}
\newcommand{\rppb}{\ensuremath{ R_{\mathrm{pPb}}}}
\begin{document}
\title{Charmonium production at mid-rapidity in Pb-Pb and p-Pb collisions with ALICE}

\author{Steffen Georg Weber on behalf of the ALICE Collaboration}

\address{Research Division, GSI Helmholtzzentrum f\"ur Schwerionenforschung, Darmstadt, Germany}

\ead{S.Weber@gsi.de}

\begin{abstract}
We  present an overview of the ALICE measurements on the production of \jpsi{} in Pb-Pb collisions at a centre-of-mass energy per nucleon pair of $\snn = 2.76\tev$ and p-Pb collisions at $\snn= 5.02\tev$ at mid-rapidity ($|y_\mathrm{lab} | < 0.8$) down to zero transverse momentum. The cold nuclear matter effects estimated from the p-Pb measurements and their impact on the interpretation of Pb-Pb results are discussed, based on comparison of data to model calculations.
\end{abstract}

\section{Introduction}

Charmonium production has been proposed as a sensitive probe of the hot and dense medium created in ultra-relativistic heavy-ion collisions. The original idea \cite{matsuisatz} was that in nucleus-nucleus collisions the binding potential between the charm and anticharm quarks is screened due to the high density of colour charges via a mechanism similar to the Debye screening. Results at SPS \cite{sps} and RHIC \cite{rhic} energies indeed showed a suppression of the \jpsi{} yield in most central collisions, beyond what was expected from cold nuclear matter (CNM) effects. 

At LHC energies, charm-anticharm quark pairs are abundantly produced in initial hard partonic processes. For this reason, it has been suggested that a new production mechanism based on the (re)combination of independently produced charm and anticharm quarks to charmonium pairs becomes relevant. These models assume either a creation of charmonium states at the  hadronization stage, like the Statistical Hadronization Model (SHM) \cite{shm} or a continuous dissociation and regeneration throughout the lifetime of the hot medium, addressed in transport models \cite{transport}.

The experimental observable used to quantify modifications of charmonium production in nucleus-nucleus collisions compared to proton-proton collisions is the nuclear modification factor, \raa, defined as 

\[
\raa = \frac{1}{ \langle T_\mathrm{AA} \rangle } \frac{\mathrm{d}^2N_\mathrm{AA}^{\jpsi} / \mathrm{d}p_\mathrm{T} \mathrm{d}y }{\mathrm{d}^2\sigma_\mathrm{pp}^{\jpsi} / \mathrm{d}p_\mathrm{T} \mathrm{d}y},
\]
where $T_\mathrm{AA}$ is the nuclear overlap function determined by Glauber model calculations.

In addition to hot medium effects, also cold nuclear matter effects can influence the production of charmonia, the most relevant ones at LHC energies are listed below. The parton distribution functions (PDF) in a nucleus are modified compared to the one in a free nucleon, this effect is known as shadowing \cite{shadowing}. There is a limited knowledge on gluon PDFs in heavy ion collisions at the moment, new data on charmonium production can provide further constraints.

\begin{figure}[t]
\begin{minipage}{17pc}
\includegraphics[width=17pc]{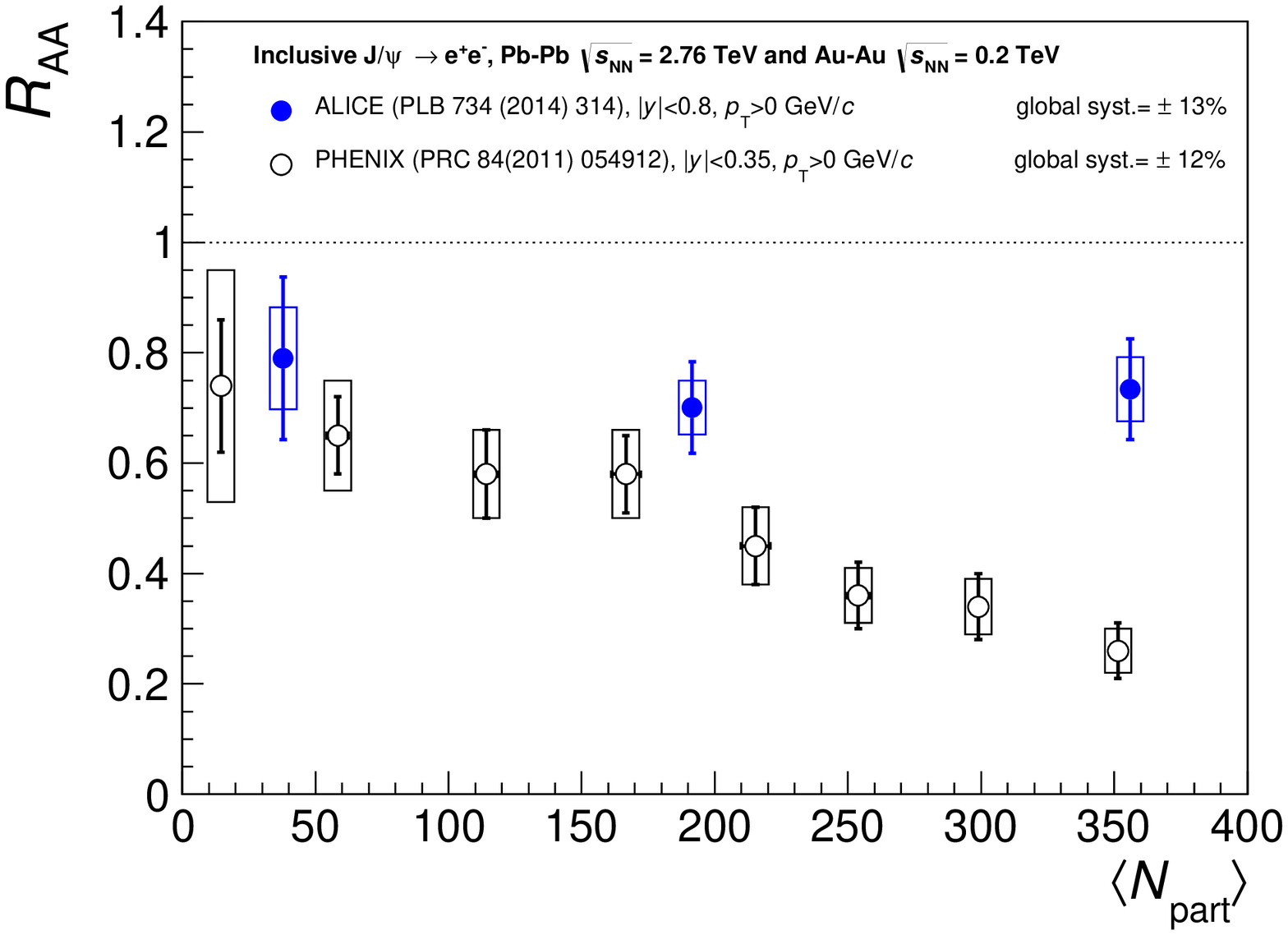}
\end{minipage}
\begin{minipage}{18pc}
\includegraphics[width=18pc]{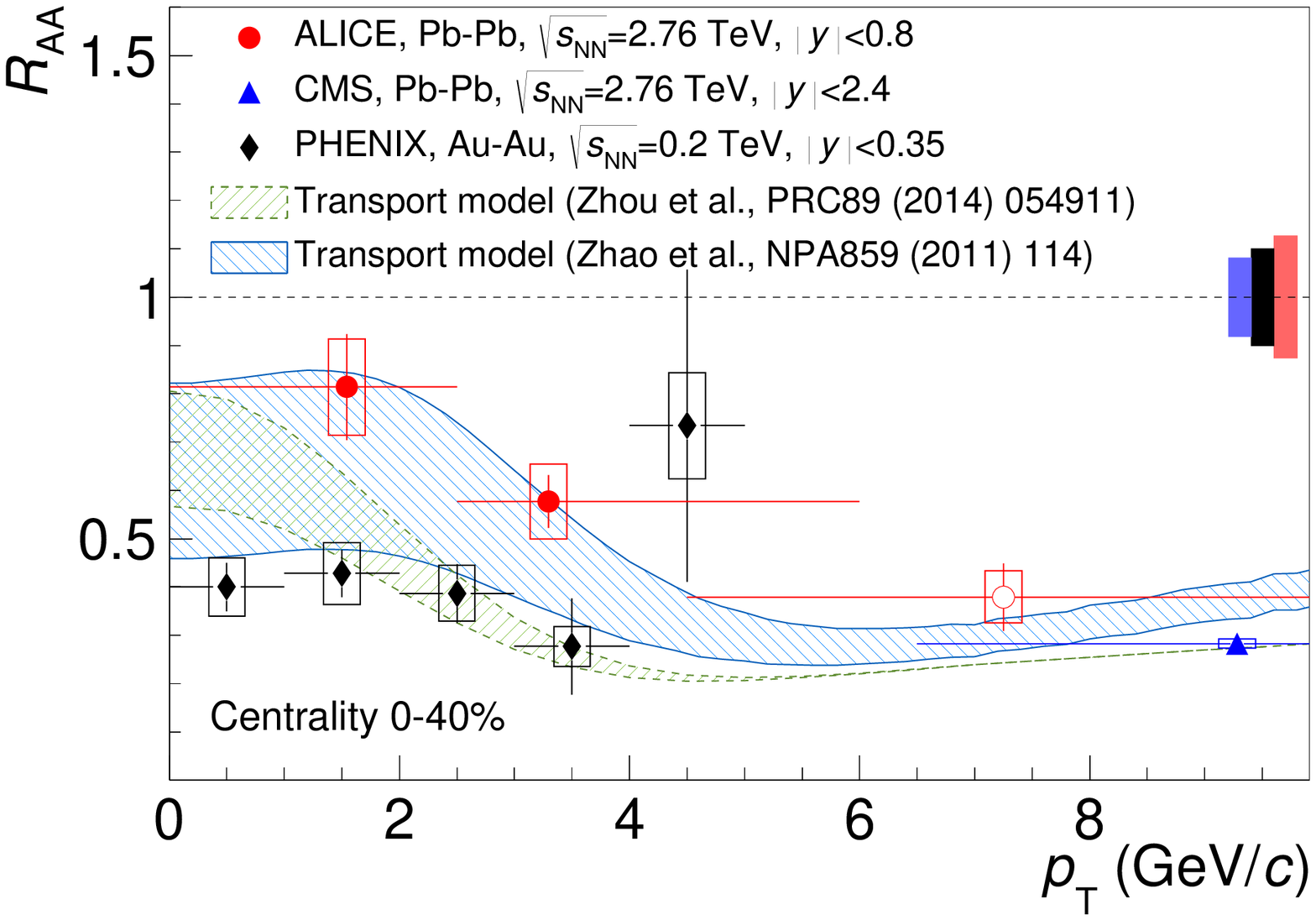}
\end{minipage} 
\caption{Left: The nuclear modification factor \raa{} measured by ALICE in Pb-Pb collisions at $\snn=2.76\tev$ as a function of the mean number of participants $\langle N_\mathrm{part} \rangle $\cite{alice_jpsi_pbpb}, compared with PHENIX \cite{rhic2} results in Au-Au collisions at $\snn=200\gev$. Right: \raa{} as function of transverse momentum \cite{alice_jpsi_pbpb_prompt}, compared with measurements from CMS \cite{cms} and PHENIX \cite{rhic} and with theoretical models \cite{transport1, transport2}.}
\end{figure}
\label{RAA}

Another way to describe the depletion of the nuclear gluon PDFs in the kinematical domain relevant for the charmonium production is via gluon saturation, which is addressed in the Color Glass Condensate (CGC) model \cite{cgc}. Other theoretical models take into account energy loss mechanisms on the partonic level, before and after the hard collision \cite{energyloss}.

The cold nuclear matter effects are experimentally accessible via proton-nucleus collisions, where the creation of a hot medium is not expected. These measurements are thus crucial for the interpretation of data from nucleus-nucleus collisions.

\section{Experimental Results}

ALICE has measured inclusive \jpsi{} production at mid-rapidity in the dielectron decay channel $\jpsi \rightarrow \mathrm{e}^+ \mathrm{e}^-$ and in forward rapidity in the dimuon decay channel $\jpsi \rightarrow \mu^+ \mu^-$. In both cases, \jpsi{} are reconstructed down to $p_\mathrm{T} = 0$. Here, measurements at mid-rapidity will be discussed. The pseudorapidity coverage for measurements in Pb-Pb collisions is $|\eta| < 0.8$, for p-Pb collisions, the coverage is shifted to $- 1.37 < \eta_\mathrm{cms} < 0.43$ in the centre-of-mass-frame due to the different per nucleon energy of the two collision partners. The main detectors which have been used for the presented results are the Inner Tracking System (ITS) for collision vertex determination and tracking, the Time Projection Chamber (TPC), for tracking and particle identification, and the V0 and Zero Degree Calorimeter (ZDC) detectors for centrality determination and event selection. 

The raw \jpsi{} yield has been obtained using the bin counting method from the invariant mass distribution of dielectron pairs in the window $2.92 < m_{\mathrm{ee}} < 3.16\,\mathrm{GeV}/c^2$ after combinatorial background subtraction using the mixed-event technique. For the calculation of the nuclear modification factor, the pp reference was obtained with an interpolation procedure including data at mid-rapidity from ALICE \cite{alicepp}, PHENIX \cite{rhic3} and CDF \cite{cdf}.

The presented results from Pb-Pb collisions are based on data taken in 2010 and 2011 at a collision energy of $\snn = 2.76\tev$ with an integrated luminosity $\mathcal{L}_\mathrm{int} = 27.7\,\mu \mathrm{b}^{-1}$, the results from p-Pb collisions are based on data from 2013 at a collision energy of $\snn = 5.02\tev$ with $\mathcal{L}_\mathrm{int} = 51.4\,\mu\mathrm{b}^{-1}$. More details of the analysis can be found in \cite{alice_jpsi_pbpb} for the Pb-Pb analysis and in \cite{alice_jpsi_ppb} for the p-Pb analysis.

\begin{figure}[t]
\begin{minipage}{17pc}
\includegraphics[width=17pc]{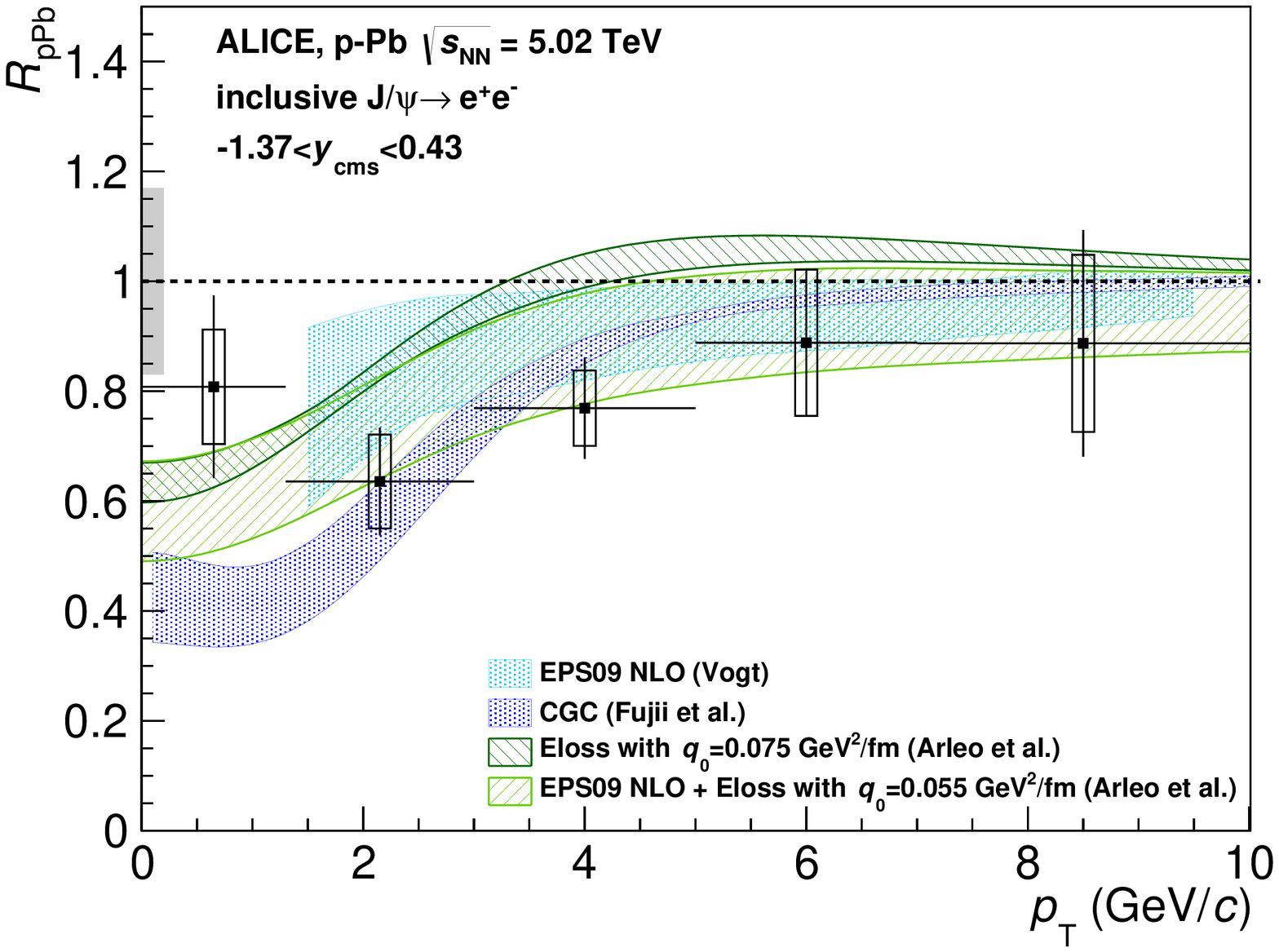}
\end{minipage}
\begin{minipage}{17pc}
\includegraphics[width=17pc]{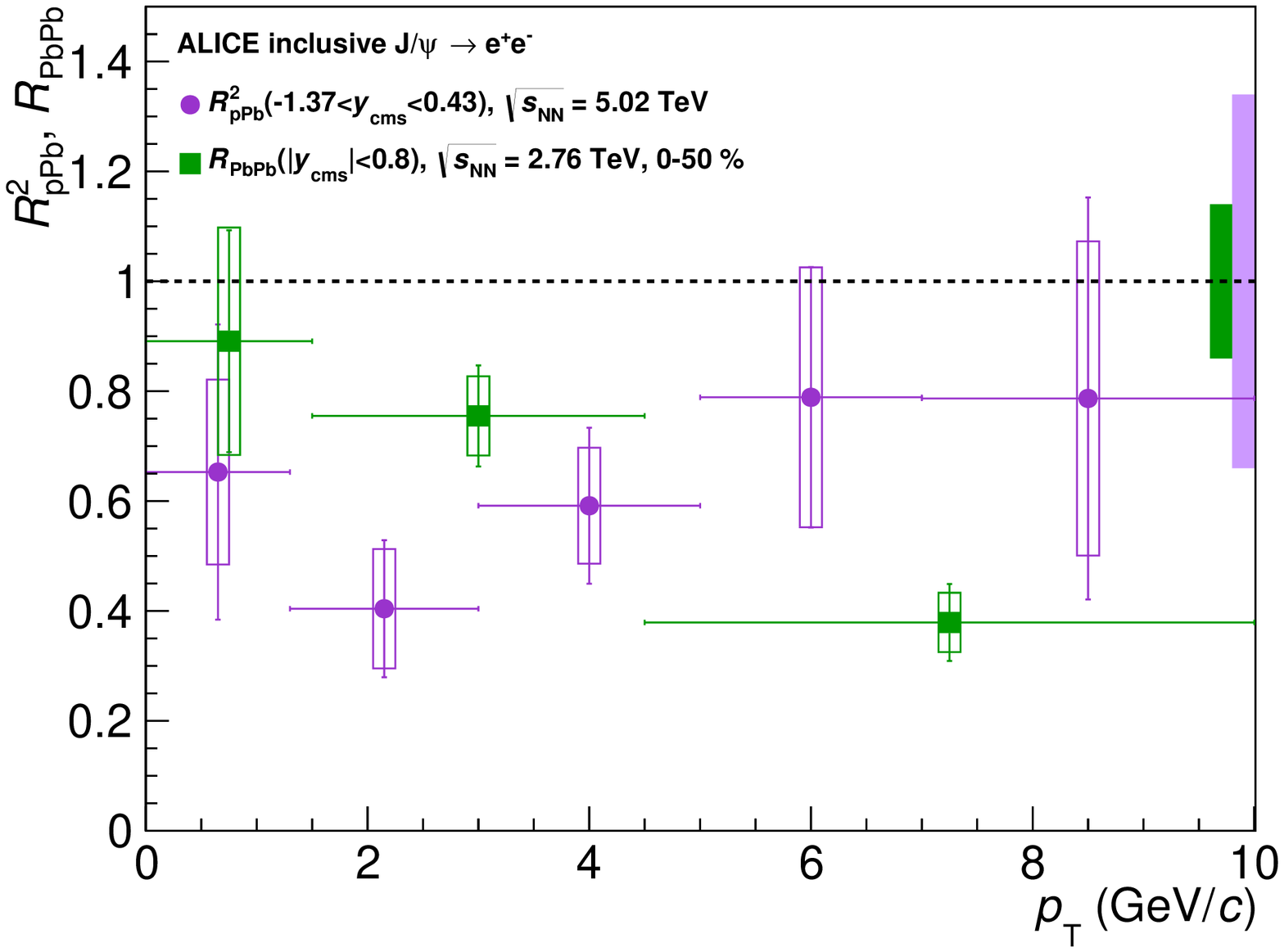}
\end{minipage} 
\caption{Left: The nuclear modification factor \rppb{} measured by ALICE in p-Pb collisions at $\snn=5.02\tev$ as a function of \pT{} \cite{alice_jpsi_ppb}, together with model calculations \cite{shadowing1, cgc1, eloss1}. Right: Squared nuclear modification factor $R_\mathrm{pPb}^2$ as an estimate for CNM effects in Pb-Pb collisions and \raa from Pb-Pb collisions at $\snn=2.76\tev$.}
\end{figure}
\label{RpPb}

The left panel of Fig. \ref{RAA} shows the \pT{}-integrated nuclear modification factor \raa{} from Pb-Pb collisions at mid-rapidity as a function of the average number of participating nucleons $\langle N_\mathrm{part} \rangle $. The ALICE results \cite{alice_jpsi_pbpb} are compared to measurements by PHENIX in Au-Au collisions at $\snn=200\gev$ \cite{rhic2}. The results for central collisions are significantly different, with the ALICE data showing a factor of three less suppression than the PHENIX results. Also the PHENIX results indicate a strong centrality dependence, with a suppression increasing with centrality, whereas the ALICE results are compatible with no centrality dependence. The same behaviour is seen at forward rapidity \cite{alice_jpsi_fwd}. The pt dependence of the \jpsi{} \raa{} for the 40\% most central collisions is shown in the right panel of Fig. \ref{RAA}. Our data is compared to PHENIX results in Au-Au collisions at $\snn=200\gev$ \cite{rhic}, CMS results at higher \pT{} \cite{cms} and model calculations \cite{transport1, transport2}. The ALICE data suggests a much lower degree of suppression at low \pT{} compared to lower energy measurements. Our data is well described by transport model predictions by Zhao et al. \cite{transport2}. Newer predictions by Zhou et al. \cite{transport1} are systematically below the measurement and exhibit a \pT{} dependence similar to the one in the data. The model calculations show substantial theoretical uncertainties due to the limited knowledge of charm cross section and cold nuclear matter effects at LHC energies. 

Another experimental observable adopted to gain further insight on charmonium production mechanisms is the mean transverse momentum $\langle \pT \rangle$ of \jpsi{} produced in different collisions systems. ALICE measurements \cite{alice_jpsi_pbpb_prompt} have shown that the $\langle \pT \rangle$ for Pb-Pb collisions is significantly smaller than that for pp collisions. Such a behaviour is not observed at smaller centre-of-mass energies from meausrements by PHENIX and NA50, for which no significant system size dependence of $\langle \pT \rangle$ is seen. This might indicate the onset of processes which either deplete the high \pT{} region or enhance the \jpsi{} production at low \pT{} in heavy-ion collisions at the LHC. The latter effect would be expected as a consequence of a significant contribution from charmonium recombination.

On the left panel of Fig. \ref{RpPb} the nuclear modification factor in p-Pb collisions \rppb{} is shown as a function of \pT, together with model calculations. These consider either pure gluon shadowing effects \cite{shadowing1}, gluon saturation \cite{cgc1} or partonic energy loss, with or without inclusion of nuclear shadowing contributions \cite{eloss1}. All considered models describe the data reasonably well.

In order to compare the results obtained in Pb-Pb and in p-Pb collisions, it is important to notice that for the investigated energies and rapidity regions, the probed Bjorken-$x$ ranges are comparable. Using 2 $\rightarrow$ 1 ($gg\rightarrow\jpsi{}$) kinematics, in Pb-Pb collisions the probed range is $7.0\times10^{-4} < x < 3.5\times10^{-3}$ and in p-Pb collisions $6.1 \times 10^{-4} < x < 3.0 \times 10^{-3}$. Under the assumption that shadowing is the main CNM-related mechanism that plays a role in the charmonium production and that its effect on the two colliding nuclei in Pb-Pb collisions can be factorized, the squared nuclear modification factor $R_\mathrm{pPb}^2$ can be considered as an estimate of CNM effects in Pb-Pb collisions \cite{cnm1, cnm2}.

The right panel of Fig. \ref{RpPb} shows $R_\mathrm{pPb}^2$ together with \raa{} as a function of \pT{}. The $R_\mathrm{pPb}^2$ shows a clear \pT{} dependence with a stronger effect at low \pT{}, whereas the \raa{} shows the largest suppression at high \pT{}. These differences suggest that the \jpsi{} suppression in Pb-Pb collisions cannot be explained by CNM effects alone. The data at low \pT{} hint at the presence of charmonium production in Pb-Pb consistent with (re)combination scenarios, whereas at high \pT{} suppression effects seem to be present in the hot medium.

\section{Conclusions}

Measurements by ALICE of inclusive \jpsi{} production at mid-rapidity as a function of centrality and transverse momentum in Pb-Pb collisions at $\snn=2.76\tev$ and in p-Pb collisions at $\snn=5.02\tev$ have been presented. Uniquely at the LHC, these measurements reach down to $\pT=0$. In Pb-Pb collisions, a suppression of \jpsi{} production is observed, with no strong centrality dependence and a weak suppression at low transverse momentum. The results are in contrast to measurements at RHIC energies, which show an increase of the suppression with centrality, especially at low \pT{}. Our data can be well described by theoretical models which take into account a (re)combination component in the \jpsi{} production. 

In p-Pb collisions, a suppression of \jpsi{} production at low transverse momenta was observed, which can be well described by models taking into account either shadowing effects, gluon saturation, or partonic energy loss. An estimate of CNM effects in Pb-Pb via the squared nuclear modification factor $R_\mathrm{pPb}^2$ in p-Pb collisions underlines the importance of hot nuclear matter effects for the interpretation of the results in Pb-Pb collisions and supports the idea of charmonium production in the hot medium or at hadronization.

\section*{References}
\bibliography{bibliography}

\providecommand{\newblock}{}
\begin{thebibliography}{10}
\expandafter\ifx\csname url\endcsname\relax
  \def\url#1{{\tt #1}}\fi
\expandafter\ifx\csname urlprefix\endcsname\relax\def\urlprefix{URL }\fi
\providecommand{\eprint}[2][]{\url{#2}}

\bibitem{matsuisatz}
Matsui T and Satz H 1986 {\em Phys. Lett.\/} B {\bf 178} 416

\bibitem{sps}
Alessandro B {\em et~al.\/} (NA50) 2005 {\em Eur. Phys. J.\/} C {\bf 39}
  335--345

\bibitem{rhic}
Adare A {\em et~al.\/} (PHENIX) 2007 {\em Phys. Rev. Lett.\/} {\bf 98} 232301

\bibitem{shm}
Braun-Munzinger P and Stachel J 2000 {\em Phys. Lett.\/} B {\bf 490} 196--202

\bibitem{transport}
Thews R~L, Schroedter M and Rafelski J 2001 {\em Phys. Rev.\/} C {\bf 63}(5)
  054905

\bibitem{shadowing}
Eskola K~J, Paukkunen H and Salgado C~A 2009 {\em JHEP\/} {\bf 04} 065

\bibitem{alice_jpsi_pbpb}
Abelev B {\em et~al.\/} (ALICE) 2014 {\em Phys. Lett.\/} B {\bf 734} 314--327

\bibitem{rhic2}
Adare A {\em et~al.\/} (PHENIX) 2011 {\em Phys. Rev.\/} C {\bf 84} 054912

\bibitem{alice_jpsi_pbpb_prompt}
Adam J {\em et~al.\/} (ALICE) 2015 {\em JHEP\/} {\bf 07} 051

\bibitem{cms}
Chatrchyan S {\em et~al.\/} (CMS) 2012 {\em JHEP\/} {\bf 05} 063

\bibitem{transport1}
Zhou K, Xu N, Xu Z and Zhuang P 2014 {\em Phys. Rev.\/} C {\bf 89} 054911

\bibitem{transport2}
Zhao X and Rapp R 2011 {\em Nucl. Phys.\/} A {\bf 859} 114--125

\bibitem{cgc}
Gelis F, Iancu E, Jalilian-Marian J and Venugopalan R 2010 {\em Ann. Rev. Nucl.
  Part. Sci.\/} {\bf 60} 463--489

\bibitem{energyloss}
Sharma R and Vitev I 2013 {\em Phys. Rev.\/} C {\bf 87} 044905

\bibitem{alicepp}
Aamodt K {\em et~al.\/} (ALICE) 2011 {\em Phys. Lett.\/} B {\bf 704} 442--455
  [Erratum: Phys. Lett.B718,692(2012)]

\bibitem{rhic3}
Adare A {\em et~al.\/} (PHENIX) 2012 {\em Phys. Rev.\/} D {\bf 85} 092004

\bibitem{cdf}
Acosta D {\em et~al.\/} (CDF) 2005 {\em Phys. Rev.\/} D {\bf 71} 032001

\bibitem{alice_jpsi_ppb}
Adam J {\em et~al.\/} (ALICE) 2015 {\em JHEP\/} {\bf 06} 055

\bibitem{shadowing1}
Albacete J, Armesto N, Baier R, Barnafoldi G, Barrette J {\em et~al.\/} 2013
  {\em Int.J.Mod.Phys.\/} E {\bf 22} 1330007

\bibitem{cgc1}
Fujii H and Watanabe K 2013 {\em Nucl.Phys.\/} A {\bf 915} 1--23

\bibitem{eloss1}
Arleo F and Peigne S 2013 {\em JHEP\/} {\bf 1303} 122

\bibitem{alice_jpsi_fwd}
Abelev B {\em et~al.\/} (ALICE) 2012 {\em Phys. Rev. Lett.\/} {\bf 109} 072301

\bibitem{cnm1}
Vogt R 2010 {\em Phys. Rev.\/} C {\bf 81}(4) 044903

\bibitem{cnm2}
Ferreiro E, Fleuret F, Lansberg J and Rakotozafindrabe A 2009 {\em Phys.
  Lett.\/} B {\bf 680} 50 -- 55

\end{thebibliography}
\end{document}